# Semantic Web Requirements through Web Mining Techniques

Hamed Hassanzadeh and Mohammad Reza Keyvanpour

*Abstract*—In recent years, Semantic web has become a topic of active research in several fields of computer science and has applied in a wide range of domains such as bioinformatics, life sciences, and knowledge management. The two fast-developing research areas semantic web and web mining can complement each other and their different techniques can be used jointly or separately to solve the issues in both areas. In addition, since shifting from current web to semantic web mainly depends on the enhancement of knowledge, web mining can play a key role in facing numerous challenges of this changing. In this paper, we analyze and classify the application of divers web mining techniques in different challenges of the semantic web in form of an analytical framework.

## I. Introduction

In a distributed informational environment, documents and objects have been joined together for convenient interactive access. The best example of such environment is World Wide Web that users utilize hyperlinks and URL addresses for finding the required information. The Web exceeds 10 billion pages, or more than six terabytes of data on about three million servers. Almost a million pages are added daily, a typical page changes in a few months, and several hundred gigabytes change every month.

But since more increasing information is imported on the web, accessing this information is getting harder than before. The main problem of this information access is because of unstructured or semi-structured web contents. So it is so difficult to structure, standardize and organize them [1]. This level of complexity in large volumes databases makes them hard to be managed or almost impossible to use information retrieval solutions in them. One of the methods to tackle this challenge is Web mining. Web mining uses data mining techniques to explore and extract information automatically from documents and web services.

Web mining tasks can be classified into three categories: Web content mining, Web structure mining, and Web usage mining. Each of these sub-categories proceeds mining of different parts of web [2].

The Semantic Web is an extension of the current web in which information is given well-defined meaning, and with changing web contents into machine understandable form, would promote quality and intelligence of the web [3].

On the other hand, because there are existed structured information and explicit metadata in the Semantic Web, the access to target information are facilitated by making it possible to semantically search the web. There are diverse challenges in Semantic Web creation. A complete classification of these challenges are introduces in [4].

As mentioned above, semantic web is not a distinct entity from the present web. On the other hand, for using data which are available in the web in an efficient way for semantic web, the level of these data must be upgraded and their embedded knowledge must be extracted before. One of the efficient methods in creating semantic web is web mining. Different web mining techniques can be used for improving data levels to information and knowledge, finding various concepts and words for building ontologies and also annotating of web data. This article presents different applications of web mining methods in semantic web field and generally the combination of these two important research fields. We introduce a framework of the role of different web mining approaches in semantic web construction tasks. This framework can give a guideline for future researches on the Semantic Web.

The rest of this paper is organized as follows: Section II briefly reviews the web mining and its methods. Section III presents the idea and general perspectives of semantic web. Section IV describes the combination of two web mining and semantic web fields and we conclude in Section VI.

## II. Web Mining

Web mining is the use of data mining techniques to automatically discover and extract information from Web documents and services [5]. In other words, Web Mining is the extraction of interesting and potentially useful patterns and implicit information from artifacts or activity related to the World Wide Web. Web mining refers to overall process of information extraction, not just the use of softwares that apply standard data mining tools. The overall process of Web Mining is decomposed into these subtasks:
1) Resource finding: the task of retrieving intended Web documents.
2) Information selection and preprocessing: automatically selecting and pre-processing specific information from retrieved Web resources.
3) Generalization: automatically discovers general patterns at individual Web sites as well as across multiple sites.
4) Analysis: validation and interpretation of the mined patterns.

Manuscript received June 2, 2012; revised July 3, 2012.
Hamed Hasszandeh is with the Department of Electronic, Computer and IT Islamic Azad University Qazvin, Iran and Young Researchers Club, Qazvin Branch, Islamic Azad University, Qazvin, Iran (e-mail: h.hassanzadeh@qiau.ac.ir).
Mohammad Reza Keyvanpour is with the Department of Computer Engineering Alzahra University Tehran, Iran (e-mail: keyvanpour@alzahra.ac.ir).

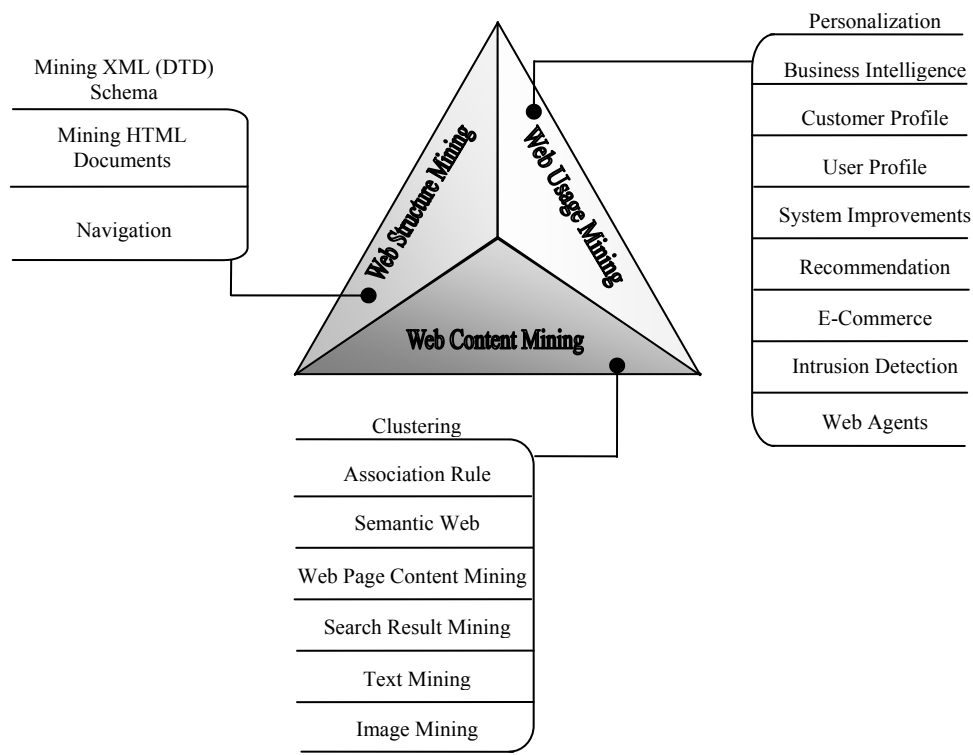

Fig. 1. Web mining techniques and applications.

There are three factors affecting the way a user perceives and valuates a site: content, Web page design, and overall site design. The first factor concerns the goods, services, or data offered by the site. The other factors concern the way in which the site makes con-tent accessible and understandable to its users. We distinguish between the design of individual pages and the overall site design, because a site is not simply a collection of pages—it is a network of related pages. The users will not engage in exploring it unless they find its structure intuitive.

Based on which part of the Web to mine, Web mining can be categorized to three areas:
1) *Web-content mining* - describes the discovery of useful information from Web documents. Basically, Web content consists of several types of data such as text, image, audio, video, metadata as well as hyperlinks. Research in mining multiple types of data is now termed multimedia-data mining. We could consider multimedia-data mining as an instance of Web-content mining. The Web content data consist of unstructured data such as free text, semi-structured data such as HTML documents, and a more structured data such as tables and database- generated HTML pages. The goal of Web-content mining is mainly to assist or to improve information-finding or filtering the information. Building a new model of data on the Web, more sophisticated queries other than the keywords-based search could be asked.
2) *Web-structure mining* - tries to discover the model underlying the link structure on the Web. The model is based on the topology of the hyperlinks with or without a description of the links. The model can be used to categorize Web pages and is useful for generating information such as the similarity relationship between Web sites.
3) *Web-usage mining* - tries to make sense of the data generated by the Web surfer's sessions or behaviors. While Web-content mining and Web-structure mining utilize real or primary data on the Web, Web-usage mining mines the secondary data derived from the behavior of users while interacting with the Web. This includes data from Web server-access logs, proxy-server logs, browser logs, user profiles, registration data, user sessions or transactions, cookies, bookmark data, and any other data that is derived from a person's interaction with the Web. Web usage mining process can be divided into three independent tasks: Preprocessing, pattern discovery and pattern analysis [5], [6].

Web mining techniques and applications are depicted in Fig. 1. Web mining typically encompasses ways of improving search or customization by (i) learning interests of users based on access patterns, (ii) providing users with pages, sites, and advertisements of interest, and (iii) using XML to improve search and information discovery on the Web [7].

### III. SEMANTIC WEB

The Semantic Web is an extension of the current web in which information is given well defined meanings that improved the interoperability between machines and human [3]. The idea of semantic web is to leave most of tasks and decisions to machines. This is applicable with adding knowledge to web contents by understandable languages for machine and establish intelligent software agents that able to process this information. On the other hand, while the Semantic Web consists of structured information and explicit metadata, it paves the way to rapidly access information and ability of semantic search [8].

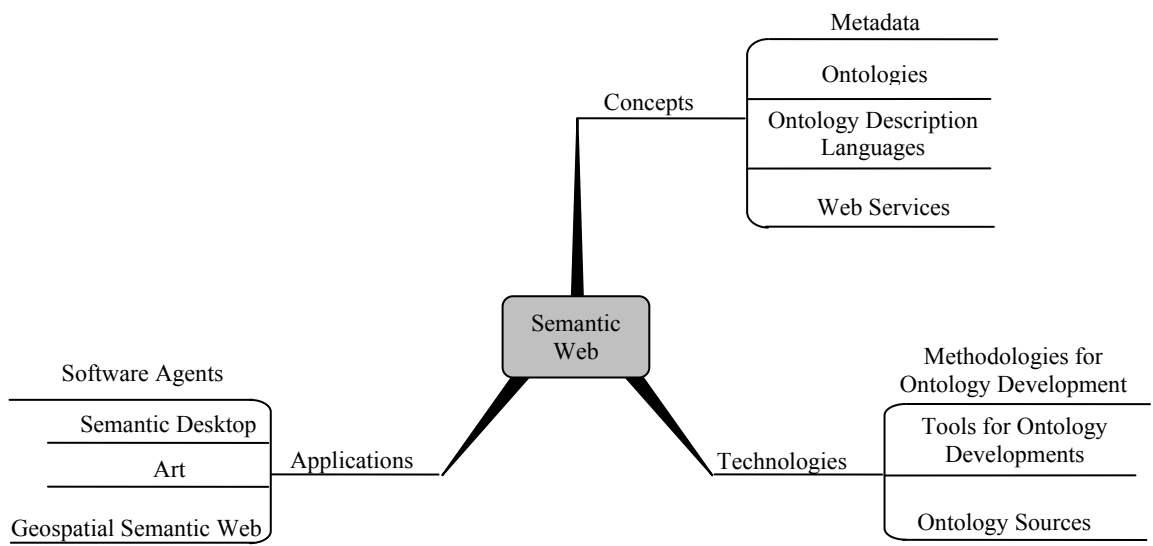

Fig. 2. Semantic web perspectives [9].

The idea of the Semantic Web is introduced by the inventor of the World Wide Web, Tim Berners-Lee. The perspective of semantic web, its technologies and applications is depicted shortly in Fig. 2 [9]. Semantic web is made of these elements:

*Uniform Resource Identifier (URI)*: A universal resource identifier is a formatted string that serves as a means of identifying abstract or physical resource. A URI can be further classified as a locator, a name, or both.

*Resource Description Framework (RDF)*: RDF contains the concept of an assertion and allows assertions about assertions. Meta-assertions make it possible to do rudimentary checks on a document. RDF is a model of statements made about resources and associated URI. Its statements have a uniform structure of three parts: subject, predicate, and object.

*Metadata*: Metadata are data about data. They serve to index Web pages and Web sites in the Semantic Web, allowing other computers to acknowledge what the Web page is about.

*Ontology*: Ontology is an agreed vocabulary that provides a set of well-founded constructs to build meaningful higher level knowledge for specifying the semantics of terminology systems in a well-defined and unambiguous manner. For a particular domain, ontology represents a richer language for providing more complex constraints on the types of resources and their properties. Compared to a taxonomy, ontologies enhance the semantics of terms by providing richer relationships between the terms of a vocabulary.

*Software Agents*: An intelligent agent is a computer system that is situated in some environment, and that is capable of autonomous action and learning in order to meet its design objectives.

*Web Services:* A Web service is a software system designed to support interoperable machine-to-machine interaction over a network. Semantic web services are built around universal standards for the interchange of semantic data, which makes it easy for programmers to combine data from different sources and services without losing meaning [10]. Fig. 3 reveals the basic requirements of semantic web [11].

## IV. SEMANTIC WEB THROUGH WEB MINING

The two fast-developing research areas Semantic Web and Web Mining build both on the success of the World Wide Web (WWW). They complement each other well because they each address one part of a new challenge posed by the great success of the current WWW: The nature of most data on the Web is so unstructured that they can only be understood by humans, but the amount of data is so huge that they can only be processed efficiently by machines. The Semantic Web addresses the first part of this challenge by trying to make the data (also) machine understandable, while Web Mining addresses the second part by (semi-) automatically extracting the useful knowledge hidden in these data, and making it available as an aggregation of manageable proportions [12].

Different aspects of web mining usage in semantic web are shown in Fig. 4. We describe these applications shortly in next subsections.

Semantic Web Mining aims at combining the two areas Semantic Web and Web Mining. This vision follows our observation that trends converge in both areas: increasing numbers of researchers work on improving the results of Web Mining by exploiting (the new) semantic structures in the Web, and make use of Web Mining techniques for building the Semantic Web.

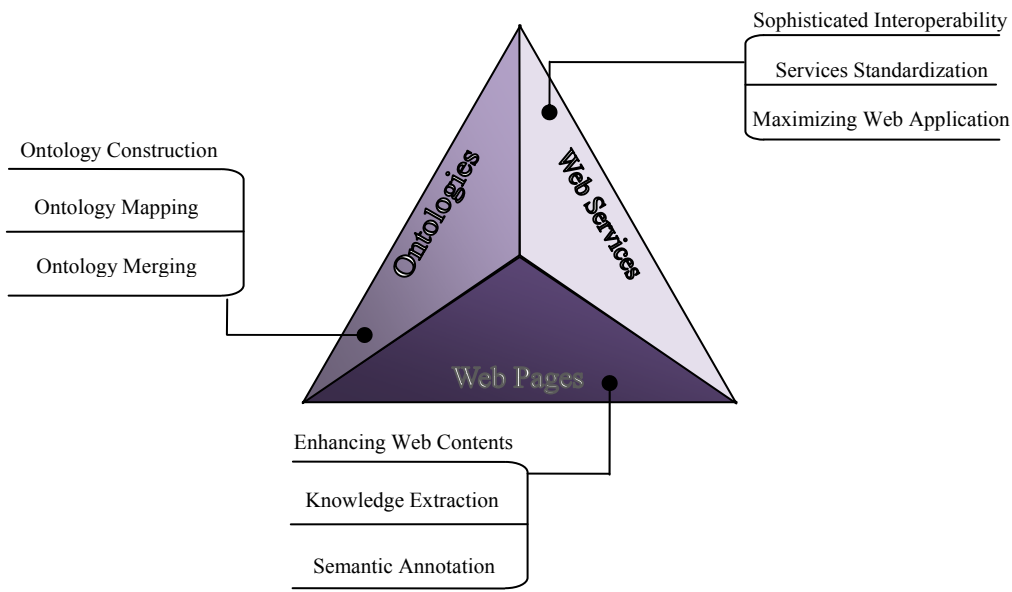

Fig. 3. Semantic web major requirements and tasks.

Last but not least, these techniques can be used for mining the Semantic Web itself. The wording Semantic Web Mining emphasizes this spectrum of possible interaction between both research areas: it can be read both as Semantic (Web Mining) and as (Semantic Web) Mining [13], [14].

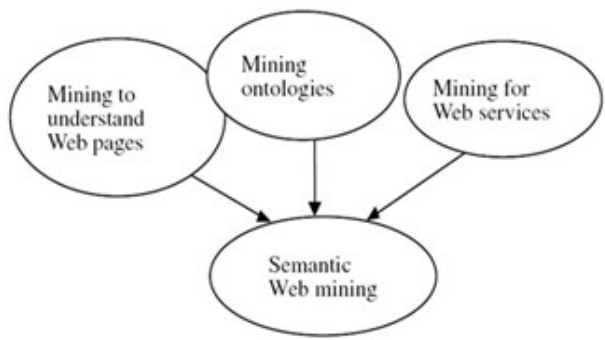

Fig. 4. Semantic web mining [8].

*A. Mining to Understand Web Pages*

One way for acquiring required knowledge for semantic web is to enhance the contents of the current web. During the years from existence of World Wide Web, a huge distributed database of different data types such as documents, images, sounds and videos is created. Undoubtedly, through the knowledge acquisition processes for semantic web, one must consider this huge database. On the other hand, the semantic level of available knowledge in current web is not enough for semantic web. So, there are needed for approaches that enhance the semantic level of knowledge in current. From the three kind of web mining, web content mining plays an important role in gathering knowledge for developing the semantic web [12].

*B. Ontology Mining*

The relation between ontologies and web mining techniques is a mutual relationship. Mining web contents can help to develop ontologies. For example, mining data in a specific domain can be applied to find related concepts in that domain and add them into domain ontology. This may help the ontology construction process [8].

Furthermore, mining the available ontologies on the web can improve the level of intelligence of the current web. Whereby ontologies are constructed from metadata, mining them besides mining the contents of the web can be applied in developing the semantic web perspectives, that is, the information that obtains from ontologies and available schemas can be used for better understanding of web pages.

*C. Web Services and Web Mining*

Web mining techniques and in general, data mining approaches not only can be used inside the web services, but also can use them to develop web services. For example, a user can use a data mining service to explore data resources on the web. Data mining can be used in intermediate services for adopting between clients and servers. Whereby web services and web mining have different components, a combination of these two techniques will leverage the potential of the Web service technology [8], [13].

## V. ANALYTICAL FRAMEWORK

This research ends in an analytical framework which is shown in Table I. As mentioned in previous sections, different web mining techniques are applied in addressing some of the Semantic Web challenges and requirements such as ontology construction, enhancing web contents, and web services. The details of these applications are shown in Table I. These results are concluded by closely studying some recent and legitimate works in this domain [12], [14]-[20].

The semantic web requirements are considered in three main groups, i.e. ontologies, web contents, and web services. Ontology construction and ontology management and improvements are important tasks that can be solved by web mining techniques. Also, different web mining methods can be applied for semantic annotation and metadata management requirements.

As revealed in Table I, the most applicable techniques of web mining for facing some of well-known semantic web challenges is web content mining. The reason is that the main focus of the semantic web is on knowledge and enhancing the informativeness level of current web contents. Other web mining techniques are rarely applied

TABLE I: ANALYTICAL FRAMEWORK OF WEB MINING APPLICATIONS FOR THE SEMANTIC WEB REQUIREMENTS

| | Ontology | | Web Contents | | Web Services |
|---|---|---|---|---|---|
| | **Ontology Construction** | **Ontology Management and Improvement** | **Semantic Annotation** | **Metadata Access and Management** | |
| **Web Content Mining** | Highly used | Highly used | Highly used | Highly used | Rarely used |
| **Web Structure Mining** | Rarely used | Not used | Rarely used | Rarely used | Not used |
| **Web Usage Mining** | Rarely used | Highly used | Highly used | Rarely used | Highly used |

## VI. CONCLUSIONS

In this paper we reviewed and analyzed the relations of two fast developing domains of Semantic Web and Web mining. At first Web mining and its well-known approaches were defined. Then a general definition of Semantic Web and its perspectives and essential components were presented. After that, within the Semantic Web requirements framework, we review the application of Web mining techniques for addressing the issues of Semantic Web implementation. As we discussed before, Web content mining has more applications among other web mining techniques, namely, Web structure mining and Web usage mining. Results are shown that different Web mining approaches can be applied for mining web pages to enhance data into knowledge, finding proper candidate for using in ontologies, mining ontologies, and mining web services. Note that, because of various challenges of Semantic Web and its wide range of research domain, it's concluded that more researches are needed specially related to web services.

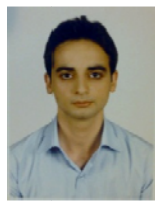

**Hamed Hassanzadeh** received his B.S. in Software Engineering from Islamic Azad University, Lahijan Branch, Lahijan, Iran. He also received his M.Sc. in Software Engineering at Islamic Azad University, Qazvin Branch, Qazvin, Iran. His research interests include Semantic Web and Machine Learning.

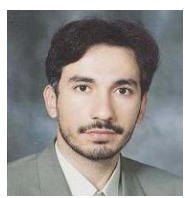

**Mohammad Reza Keyvanpour** is an Assistant Professor at Alzahra University, Tehran, Iran. He received his B.S. in Software Engineering from Iran University of Science andTechnology, Tehran, Iran. He received his M.S. and Ph.D. in Software Engineering from Tarbiat Modares University, Tehran, Iran. His research interests include image retrieval and data mining.